\documentclass[11pt,twoside]{article}

\usepackage{asp2006}
\usepackage{epsf}
\usepackage{epsfig}
\usepackage{lscape}

\begin{document}
\title{Galactic disc warps due to intergalactic accretion flows onto the disc}   
\author{M. L\'opez-Corredoira$^1$, J. Betancort-Rijo$^{1,2}$, J. E.
Beckman$^{1,3}$}   
\affil{$^1$ Instituto de Astrof\'\i sica de Canarias, La 
Laguna, Tenerife, Spain\\
$^2$ Departamento de Astrof\'\i sica, Univ. La Laguna, Tenerife,
Spain\\
$^3$ Consejo Superior de Investigaciones Cient\'\i ficas (CSIC), Spain}    

\begin{abstract} 
The accretion of the intergalactic medium onto the gaseous 
disc is used to explain
the generation of galactic warps. A cup-shaped distortion
is expected, due to the transmission of the linear momentum; but, 
this effect is small for most incident inflow angles 
and the predominant effect turns 
out to be the transmission of angular momentum, i.e. a torque giving an 
integral-sign shaped warp. 
The torque produced by a flow of velocity 
$\sim 100$ km/s and baryon density $\sim 10^{-25}$ kg/m$^3$, which is 
within the possible values for the intergalactic medium, is enough to 
generate the observed warps and this mechanism offers quite a plausible 
explanation. The inferred rate of
infall of matter, $\sim 1$ M$_\odot$/yr, to the Galactic disc that this theory
predicts agrees with the quantitative predictions 
of chemical evolution resolving key issues, notably
the G-dwarf problem.

S\'anchez-Salcedo (2006) suggests that 
this mechanism is not plausible because it
would produce a dependence of the scaleheight of the disc 
with the Galactocentric azimuth in the outer disc, but rather than being 
an objection this is another argument in favour of the mechanism
because this dependence is actually observed in our Galaxy.

\end{abstract}



\section{The mechanism}

In L\'opez-Corredoira et al. (2002, hereafter LBB), we proposed 
a mechanism to explain the formation
of warps (S-shaped or U-shaped or a combination of both)
in spiral galaxies: 
the accretion of the intergalactic medium onto the disc.
In a Milky-Way-like galaxy, the mean density of baryonic matter in 
the intergalactic medium needed to produce the observed warp 
is around $10^{-25}$ kg/m$^3$ when the 
infall velocity at large distance is $\sim 100$ km/s (LBB).
These numbers were confimed independently by S\'anchez-Salcedo (2006).
This hypothetical low density net flow is a very reasonable physical assumption
and would explain why most spiral galaxies are warped.
No massive halo is necessary nor are high values of magnetic fields nor
satellite companions are necessary, and the presence 
of these elements would not modify qualitatively the present conclusions. 

\section{Why accretion onto the disc rather than into the halo?}

If the halo axis were misaligned with respect to the disc axis due
to the accretion of intergalactic matter into the disc, the disc 
would be torqued by the halo and this 
will give rise to warps too (Ostriker \& Binney 1989).
However, our opinion is that our LBB mechanism is preferable. 
The direct accretion of matter by the disc is more plausible than the accretion 
by the halo because the low density of baryonic matter in the halo produces
a very small friction to trap intergalactic flows. 
Moreover, there is no correlation between
the amplitude of the warps and the mass of the halos (derived from
rotation curves) (Castro-Rodr\'\i guez et al. 2002). 
The lenticular galaxies, which have discs with no gas, do not show warps 
(S\'anchez-Saavedra et al. 2003). If the mechanism to produce
warped discs were a purely gravitational interaction, such as the interaction
halo-disc proposed by Ostriker \& Binney (1989), it would not distinguish
between gas and stars and lenticular might also have a warp. 
However, it seems that the gas in the disc
plays an important role, in favour of our mechanism which requires
the friction with the disc gas to trap the intergalactic matter, although
other explanations in terms of lower mass of the halo in lenticulars
are in fact possible.

Anyway, even if the accretion onto the halo were a possible mechanism,
the continuous accretion of low metal gas onto the disc is also a fact
(Tinsley 1980), so the mechanism proposed here can always act.

\begin{figure}
\begin{center}
\mbox{\epsfig{file=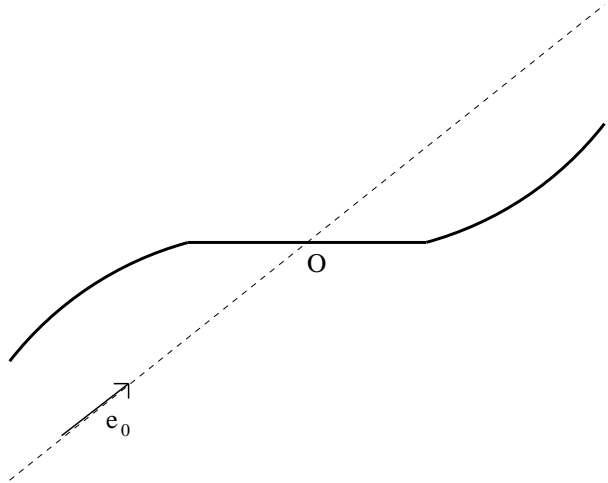,width=8cm}}
\end{center}
\caption{Graphical representation of the orientation of the S-warp with
respect to the velocity of incoming flow. The azimuthal angle $\phi _W$ 
with maximum amplitude of the warp, which is not plotted, 
is the same as the azimuthal angle of $\vec{e_0}$.}
\label{Fig:warpdirec}
\end{figure}

\section{Different types of observations 
successfully explained by this hypothesis}

The accretion of intergalactic flows with these characteristics
gives rise to:

\begin{itemize}

\item Good agreement with the observations
of the chemical evolution of the Milky Way by contributing
$\sim 1$ M$_\odot$/yr of low metallicity gas to the disc (LBB).
This turns out to be of the order of the accretion rate required to resolve 
the G-dwarf problem in our Galaxy (Tinsley 1980)
as well as explaining a number of phenomena
of chemical evolution
which require the long-term infall of low metallicity gas.

\item The frequency of warps and its amplitude
is dependent on environment (Garc\'\i a-Ruiz et al. 2002). 
The most isolated galaxies are more frequently warped,
although with less amplitude.
It seems clear that warping is due to something related to
the environment rather than to the intrinsic properties of the galaxies,
to something which is not related with the proximity of other galaxies.
The accretion of intergalactic matter onto the disc seems a good
candidate to explain these observational facts.

\item An older version of the present mechanism was proposed
by Kahn \& Woltjer (1959), but this was rejected because it predicted only
cup-shaped (U-shape) warps. This is not a valid criticism 
anymore to the present
model. Integral-shape (S-shape) warps with the amplitude and shape
observed for instance in the Milky Way are reproduced by this theory
as a consequence of the transmission of angular momentum (LBB). Even
if the global angular momentum of the intergalactic medium is null, 
the redistribution of momentum in different rings of the galaxy
due to the gravity produces a net torque in each ring (LBB).
Figure \ref{Fig:warpdirec} illustrates the relationship between
the wind velocity and the S-shape warp.

\item  Rather than being a criticism, the fact that this scenario produces
U-shaped warps is now an argument in its favour.
U-shaped warps are explained by this mechanism
due to a transmission of linear momentum (LBB), while the S-shaped
warps are due to the transmission of angular momentum. LBB predict
that the frequency of U-shaped warps should be lower than S-warps;
only occurring when the flow 
is accreted from nearly polar directions (low angle
with the rotational axis of the disc). 
This was corroborated with a different calculation method 
by S\'anchez Salcedo (2006).
This lower frequency of U-warps over S-warps
is observed (Garc\'\i a-Ruiz et al. 2002, 
S\'anchez-Saavedra et al. 2003). 
The asymmetric cases can also be explained by the present theory as a 
combination of S-warps and U-warps (LBB; Saha \& Jog 2006).
Up to now there has not been any alternative explanation for the U-warps
or the asymmetric warps.

\item S\'anchez-Salcedo (2006) raises the criticism that 
this mechanism is not plausible because it
would produce a dependence of the scaleheight of the disc 
with the Galactocentric azimuth ($\phi $; defined to be zero in the
line Sun-galactic center) in the outer disc. 
Rather than being 
an objection, however, it is another argument in favour of our model 
because this dependence is actually observed in our Galaxy: Voskes \& Burton (2006,
Fig. 15) and Levine et al. (2006, Fig. 5) have shown that the
scaleheight of the outer disc ($R>20$ kpc) is $2-3$ times
higher on average for $0<\phi <180^\circ$ than for 
$180^\circ <\phi <360^\circ $. This is
in agreement
with the expectations of our model: the wind comes from the direction
of the southern warp, $\phi \approx 270^\circ $
[to produce U-warp northwards which causes
an smaller amplitude of the (S+U)-warp of the southern warp], 
so a higher pressure is expected for the region
around the southern warp and consequently a lower thickness therein.
This effect is expected to appear in the outer disc because here the
pressure dominates over the self-gravitation of the disc.
S\'anchez-Salcedo (2006) derived roughly a factor of 6 with our LBB
mechanism between the minimum and maximum scaleheight at $R=22$ kpc.
There are differences as high as this in Levine et al. (2006, Fig. 4),
but this factor might be somewhat lower for other reasons.
For instance, it we take into
account the response time ($t_{response}$) of the disc and the rotation 
($\omega (R)$): scaleheight corresponds to the average pressure
between $[\phi -\omega (R)t_{response}]$ and $\phi $. The fact that
the intergalactic medium might not be continuous but with some clouds
would change the factor. Also the variation of some
free and unknown parameters such as the angle of the intergalactic flow with
respect to the rotation axis of the galaxy ($\theta _0$ in LBB) produce
variations in the factor.

\end{itemize}

\section{Conclusions}

Several mechanisms can generate
warps: intergalactic magnetic fields (Battaner et al. 1990),
gravitational interaction with satellites, or with the halo
misaligned by an accretion of intergalactic matter into it (Ostriker
\& Binney 1989); or our proposed 
mechanism in LBB of accretion of intergalactic flows 
onto the disc. This last option seems
to offer a very plausible scenario: it is quantitatively 
consistent with many observations
and works independently of other ingredients of galaxies
and their structure. A key advantage of the LBB
theory is that it explains facts
like U-shaped warps or variations of scaleheight with azimuth
which other theories do not explain.

\acknowledgements 



\begin{thebibliography}{}

\bibitem{} Battaner E., Florido E., S\'anchez-Saavedra M. L., 1990,
A\&A 236, 1

\bibitem{} Castro-Rodr\'\i guez, N., L\'opez-Corredoira, M., 
S\'anchez-Saavedra, M. L., \& Battaner, E. 2002, A\&A, 391, 519

\bibitem{} Garc\'\i a-Ruiz, I., Sancisi, R., \& Kuijken, K. 2002,
A\&A 394, 769

\bibitem{} Levine, E. S., Blitz, L., \& Heiles, C. 2006, ApJ 643, 881

\bibitem{} L\'opez-Corredoira, M., Betancort-Rijo, J., \&
Beckman, J. E. 2002, A\&A, 386, 169 [LBB]

\bibitem{} Saha, K., \& Jog, C. J., 2006, A\&A, 446, 897

\bibitem{} S\'anchez-Saavedra, M. L., Battaner, E., Guijarro, A.,
L\'opez-Corredoira, M., \& Castro-Rodr\'\i guez, N. 2003, A\&A 399, 457

\bibitem{} S\'anchez-Salcedo, F. J. 2006, MNRAS, 365, 555

\bibitem{} Ostriker, E. C., \& Binney, J. J. 1989, MNRAS 237, 785

\bibitem{} Tinsley B. M., 1980, Fund. Cosmic Physics 5, 287

\bibitem{} Voskes, T., \& Burton, W. B. 2006, astro-ph/0601653

\end{thebibliography}
\end{document}